# Transcript length mediates developmental timing of gene expression across *Drosophila*

Carlo G. Artieri and Hunter B. Fraser*

Department of Biology, Stanford University, Stanford, CA 94305, USA.

***Corresponding author**

**CONTACT INFORMATION:**

Hunter B. Fraser
Herrin Labs Rm 305
371 Serra Mall
Stanford, CA 94305
United States

**TELEPHONE NUMBER:** 650-723-1849

**FAX NUMBER:** 650-724-4980

**EMAIL:** hbfraser@stanford.edu

**RUNNING HEAD:** Transcript Length and Developmental Timing

**WORD COUNT:** 5,184

**NUMBER OF FIGURES:** 5

**NUMBER OF TABLES:** 1

**NUMBER OF SUPPLEMENTS:** 1 Document, 2 Tables

**ABSTRACT**

The time required to transcribe genes with long primary transcripts may limit their ability to be expressed in cells with short mitotic cycles, a phenomenon termed intron delay. As such short cycles are a hallmark of the earliest stages of insect development, we used *Drosophila* developmental timecourse expression data to test whether intron delay affects gene expression genome-wide, and to determine its consequences for the evolution of gene structure. We find that long zygotically expressed, but not maternally deposited, genes show substantial delay in expression relative to their shorter counterparts and that this delay persists over a substantial portion of the ~24 hours of embryogenesis. Patterns of RNA-seq coverage from the 5′ and 3′ ends of transcripts show that this delay is consistent with their inability to terminate transcription, but not with transcriptional initiation-based regulatory control. Highly expressed zygotic genes are subject to purifying selection to maintain compact transcribed regions, allowing conservation of embryonic expression patterns across the *Drosophila* phylogeny. We propose that intron delay is an underappreciated physical mechanism affecting both patterns of expression as well as gene structure of many genes across *Drosophila.*

## AUTHOR SUMMARY

The transcription of genes with long introns can take minutes to hours and must finish before cells divide, since incomplete transcripts are targeted for degradation. It is known that some cell division cycles, such as those involved in early insect embryogenesis, can occur in under 10 minutes, potentially limiting the expression of long genes. We explored patterns of expression of genes in *Drosophila melanogaster* over the course of embryogenesis and found that long, but not short genes are indeed prevented from being expressed early in development. Furthermore, these long transcripts require several hours to reach stable levels of expression, revealing an underappreciated mechanism, intron delay, which limits the production of long transcripts over approximately half of fly embryogenesis. Additional data confirmed that this pattern cannot be explained by delayed transcriptional activation. We also show that this pattern is conserved across millions of years of evolution, and found evidence that short genes that are able to escape delayed expression are under substantial pressure to maintain their compact lengths. Therefore intron delay also appears to be a source of significant evolutionary constraint on how gene structures can evolve.

## INTRODUCTION

The expression of genes with long primary transcripts is likely to impose significant organismal costs, as the time required to transcribe through long introns is non-trivial [1], [2]. As an extreme example, the largest known gene, human dystrophin, requires ~16 hours for the transcription of its ~2.3 Mb primary transcript. This precludes its ability to be rapidly induced by purely transcriptional means [3]. At the level of the transcriptome, the burden of transcriptional time has manifested itself in the observation that genes with expression patterns that change rapidly in response to stress have significantly lower intron densities as compared to the genomic average [4]. Similarly, transcriptional time has also been shown to limit expression of genes with long primary transcripts in cells undergoing rapid mitotic cycles, a phenomenon termed 'intron delay' [1].

A variety of studies have shown that transcription from all three RNA polymerases ceases once cells leave interphase and enter mitotic divisions [5]. Though the precise mechanisms of this repression remain poorly understood, an important component involves lack of access of the polymerase to condensing chromatin. As the mitotic cycle begins, incomplete transcripts are released from the condensing chromosomes and are subsequently degraded by an unidentified nuclear mechanism. These transcripts remain undetectable until the completion of mitosis [6]. Accordingly, strong selection is hypothesized to exist against the expansion of existing introns (or the introduction of new introns) in genes that must be expressed in cells undergoing frequent mitoses. This is supported by the observation that single-celled eukaryotes with rapid reproductive rates, such as yeasts and *Guillardia*, have very intron-poor genomes despite having descended from more intron-rich ancestors [7], [8].

In metazoans, certain cell types and/or developmental stages have mitotic cycles rapid enough to limit the expression of long genes. For example, in flies the earliest stages of development are characterized by rapid mitotic cycles as short as 8.6 minutes per division [9]. It is during this period of development that transcription of the zygotic genome begins, a process known as zygotic genome activation. Most insects achieve these rapid mitotic cycles by avoiding cytokinesis altogether and generating nuclei within a common embryonic cytoplasm known as a syncytial blastoderm [10]. In *Drosophila*, the zygotic nucleus undergoes 13 synchronous mitotic divisions, the first 9 requiring approximately 9 min each, while cycles 10-13 progressively lengthen to a maximum of 17 min per division [9]. Subsequently, the nuclei undergo a ~60 min extended 14$^{th}$ mitotic cycle during which cellularization takes place [11], [12]. While the majority of zygotic transcription is delayed until the extended 14$^{th}$ stage, it has been shown that some genes are transcribed during the mitotic divisions. Indeed, one gene with a primary transcript length exceeding 20 kb produced aborted transcripts during these early cycles that were not exported from the nucleus and degraded gradually, supporting the predictions of the intron delay hypothesis [6], [13], [14].

Delayed expression of long transcripts may play a functional, regulatory role during early development [1], [2]. For instance, the early stages of embryogenesis in *Drosophila* involve sequential activation of very short pair-rule genes followed by significantly longer homeodomain box (HOX) genes [1]. A regulatory mechanism based solely on physical constraint is appealing as it allows for a simple sequential process of activation during early development as cell cycles lengthen, without the need to invoke more complex temporal regulatory networks [15]. Furthermore, it could also regulate spatial patterning of gene expression during later periods of embryogenesis, when the embryo is partitioned into discrete mitotic domains, the cells of which may replicate at increased rates via endocycling [16], [17].

We show that early developmental intron delay of zygotically expressed genes is a general feature of the fruit fly transcriptome, limiting the expression levels of long transcripts well into embryogenesis. Furthermore, we confirm that the expression patterns observed are not simply due to regulation of transcription initiation. Finally, we extend our observations across the *Drosophila* phylogeny and show that intron delay may impose significant selective pressure to maintain compact primary transcripts among highly expressed zygotic genes.

**RESULTS**

*Long zygotic transcripts show delayed activation during* D. melanogaster *embryogenesis*

In order to explore the relationship between transcript length and patterns of expression over the course of embryonic development of *D. melanogaster*, we obtained data from two RNA-Seq timecourses generated from poly-A selected RNA: 1) the MODel organism ENCyclopedia Of DNA Elements (modENCODE) *D. melanogaster* developmental timecourse [18], which consists of 12 sequential two hour time-synchronized developmental time points spanning the ~24 hour period of fly embryogenesis (hereafter the 'embryonic' time course), and 2) the dataset of Lott et al. [19], which consists of single embryo samples spanning syncytial cycles 10 to 13 and four time points spanning the extended 14th cycle (labeled A-D) (hereafter the 'syncytial' time course). The entire syncytial timecourse takes place during the first and second time points of the embryonic timecourse [20] (Figure 1). Expression at the gene level was calculated in Reads Per Kilobase per Million mapped reads (RPKM) (see Methods) (Table S1 contains all analyzed data). Complete zygotic genome activation does not begin until ~80 min post egg laying (hereafter, all times are indicated as post egg laying), and thus most mRNA present in the embryo prior to this time is maternally deposited. Most maternal transcripts are eliminated by ~180 min, prior to which time the zygote contains both maternal and zygotic transcripts [21]. In order to analyze transcripts derived from maternal or zygotic origins separately, we used the classifications provided by Tadros et al. [22], resulting in classifications of either 'maternal' or 'zygotic' for 7,452 genes expressed in the embryonic timecourse, and 5,644 genes in the syncytial timecourse (note that the classifications refers only to the origin of these transcripts in the embryo; once maternally deposited transcripts are eliminated, all embryonic transcripts are produced from the zygotic genome).

The intron delay hypothesis predicts that the short mitotic cycles occurring during early fly embryogenesis will not allow sufficient time for the transcription of long transcripts. Therefore, we investigated the relationship between primary transcript length and expression by binning genes in both timecourses into two categories: those with 'short' transcripts < 5 kilobases in length and 'long' transcripts ≥ 5 kb (Table 1) (see Methods). We found that zygotic transcripts are significantly shorter than those maternally deposited (embryonic timecourse median lengths with bootstrapped 95% confidence intervals were 2,287 ± 100 and 3,175 ± 109 bp respectively; Kruskal-Wallis $p < 2.2 \times 10^{-16}$; data are qualitatively similar for the syncytial timecourse). Consistent with this, there is a significant over-representation of intronless zygotic genes in the syncytial timecourse (11.1% vs. 7.8% for zygotic and maternal genes, respectively; $\chi^2 = 11.36$, 1 degree of freedom [df], $p = 0.0008$), but not in the embryonic timecourse (8.5 vs. 7.5% for zygotic and maternal genes, respectively; $\chi^2 = 1.84$, 1 df, $p = 0.175$), suggesting that introns are underrepresented only in zygotic genes expressed during the earliest stages of development, and not among all zygotic genes.

The intron delay hypothesis also predicts that the difference in expression level between short and long genes should be largest during the earliest stages of development and decrease as cell-cycle intervals lengthen during development. To test this, we performed linear regressions on the median expression levels of the two length categories of zygotic and maternal genes (which is not affected by the general tendency for higher expression of short genes [23], [24]). Although the expression of both short and long zygotic genes increases during embryogenesis (Figure 2A), the slope is twice as large for the long genes ($m = 0.502$ and 1.04, $R^2 = 0.870$ and 0.858, $p = 2.44 \times 10^{-5}$ and $3.79 \times 10^{-5}$ for short and long genes, respectively). Consistent with this, analysis of covariance (ANCOVA) revealed that the difference in expression between long and short transcripts decreases over development ($F_{3,8}$ = 30.3, p = 1.02 × $10^{-4}$) (these conclusions remain robust to the removal of any single timepoint; data not shown). In contrast to zygotic genes, maternal genes showed a completely different pattern (Figure 2B): short maternal transcripts decrease in abundance (m = - 0.911, $R^2$ = 0.903, p = 5.58 × $10^{-6}$), while long maternal genes showed no significant pattern of change (m = 0.183, $R^2$ = 0.368, p = 0.0862). This indicates that the patterns observed among zygotic genes are not a general pattern related to transcript length, but rather reflect the transcriptional dynamics of transcripts expressed from the zygotic genome.

We observed the same general patterns in the syncytial timecourse (Figure 2C-D): median expression levels of both zygotic size classes increased (m = 1.31 and 2.28, $R^2$ = 0.905 and 0.968, p = 6.88 × $10^{-4}$ and = 2.61 × $10^{-5}$, for short and long genes respectively) and ANCOVA again revealed that the difference in median expression levels between short and long genes decreased over the timecourse ($F_{3,4}$ =196.9, p = 8.47 × $10^{-5}$). In this case, both short and long maternal transcripts showed a significant decrease in median expression level (slope = -1.03 and -0.789, $R^2$ = 0.903 and 0.712, p = 5.58 × $10^{-6}$ and 0.0209, for short and long genes respectively). Therefore, both data sets support the pattern predicted by the intron delay hypothesis among genes expressed from the zygotic genome.

*RNA-Seq coverage patterns are consistent with intron delay*

Although the decreasing difference in median expression levels of short and long genes during development is consistent with the intron delay hypothesis, it could also be explained by differences in transcription initiation. However, we find no evidence of differential transcription initiation between short and long zygotic genes based on ChIP-Seq profiles of well-studied chromatin marks associated with transcriptional activation or repression over a comparable embryonic timecourse [25] (Document S1, Figures S4-5).

A key prediction unique to intron delay is the presence of incomplete transcripts. This could be tested by comparing RNA-Seq reads derived from the 5′ vs. 3′ ends of transcripts in long vs. short genes, since aborted transcripts should often lack 3′ ends. Furthermore, this ratio would be expected to decrease over time as cell cycles lengthened, allowing complete transcription of progressively longer zygotic transcripts [13]. Conversely, if the expression patterns observed were entirely the result of a widespread delay in transcriptional initiation, a relatively constant 5′:3′ ratio would be expected over the course of embryonic development (Figure 3A) (we discuss and reject a third mechanism, a kinetic model explaining the delay of long genes, in Document S1).

In order to differentiate between these two possibilities, we obtained another modENCODE RNA-Seq timecourse dataset consisting of non-poly-A selected (and therefore not 3′ biased) RNA extracted from the same 12 time points as the embryonic timecourse [18] (see Methods). We calculated RPKMs for the 5′-most 1 kb of exonic transcript as well as the 3′-most 1 kb of exonic transcript and plotted the medians of the 5′:3′ ratios at each timepoint (Figure 3 B,C). Across the first six stages (0-12 h, during which zygotic activation takes place) (Document S1), only long zygotic genes show a significant change, with the 5′:3′ ratio decreasing over time (triangles in Figure 3B: $m = -0.162$, $R^2 = 0.811$, $p = 0.00905$; $p > 0.05$ for all other categories). Extension of the regressions to all 12 time points results in a significant negative slope among all four gene categories ($p < 0.05$); however long zygotic genes show a significantly steeper negative slope than the other gene categories (ANCOVA, $p < 0.001$), as expected by predictions of the intron delay hypothesis. Median levels of exonic coverage, normalized for overall gene expression level, across the first 10 kb of transcript length show a more negative slope in zygotic as compared to maternal genes during early embryogenesis, indicating that the patterns observed in the 5′:3′ ratio are not simply an artifact of analyzing only the ends of transcripts (Document S1).

*Intron delay is observed across the* Drosophila *phylogeny*

Having identified a widespread role for intron delay in *D. melanogaster*, we sought to determine if these patterns were shared in other species of fruit fly, and whether intron delay had consequences for the evolution of gene structure or expression. We therefore analyzed a microarray timecourse spanning two-hour intervals over the first 18 h of embryonic development in six *Drosophila* species (hereafter the 'species timecourse') (Figure 1) [26]. We focused our analysis on four species with high-quality annotations: *D. melanogaster*, *D. ananassae* (~12 million years [my] divergence time from *D. melanogaster*), *D. pseudoobscura* (~45 my), and *D. virilis* (~63 my). Among the transcripts represented in the dataset, 2,067 genes were represented in the other timecourses and had identifiable 1:1 orthologs among all four species [27] (see Methods). Because significant changes in transcript lengths between species are likely to occur via changes in intron lengths—and due to the difficulty in annotating untranslated regions in these other species—we classified genes based on the length of orthologous introns within orthologous genes (see Methods): genes in each species whose orthologous intron length was $< 5$ kb were classified as short, and those with intron length $\geq 5$ kb were classified as long.

Analysis of the microarray data using the same methods as for the RNA-Seq data showed parallel results in all four species: long zygotic genes increase significantly in expression level over the timecourse ($p < 0.001$ in all cases) (Figure 4). Furthermore, the rate of increase in expression level was significantly greater for long as compared to short zygotic genes in all four species (ANCOVA: $F_{3,5}$ $p < 0.05$). Conversely, no significant trend among the median expression levels across time points was observed for short zygotic transcripts in any species after correction for multiple tests. In contrast, short maternal transcripts showed a significant decrease in median expression level in all species ($p < 0.05$) except in *D. ananassae* ($p = 0.0661$), while no significant trend was observed over the timecourse among long maternal genes in any species. Therefore, despite the lack of obvious differences among the functions of genes deposited maternally vs. expressed zygotically (see Document S1), the high degree of concordance of expression patterns among species suggests that the mode of delivery of these transcripts to the embryonic transcriptome may be largely conserved across *Drosophila*.

*Conservation of short introns in highly expressed zygotic genes*

Our observation that transcripts with longer lengths are associated with delayed embryonic expression led us to predict that zygotic genes that are highly expressed during early embryogenesis across species should be subject to selection against intron expansion. Consequently, highly expressed zygotic genes should be more conserved for short transcript lengths than other gene categories. We tested this prediction by dividing genes based on their expression levels in the first time point (0-2 h) of the species timecourse: zygotic genes in the highest and lowest-expressed quartiles across all 4 species (‘high’ and ‘low expression zygotic’; 100 and 73 genes, respectively), as well as maternal genes in these same quartiles (‘high’ and ‘low expression maternal’; 142 and 189 genes, respectively). We then asked whether orthologous intron lengths in any of the categories were more variable across the *Drosophila* phylogeny by calculating the corrected coefficient of variation (CV*) of intron lengths for the four species (see Methods) (Figure 5). The CV* values, as well as intron lengths, of highly expressed zygotic genes are significantly lower than all other categories ($p < 0.01$) (Document S1, Figure S6). This suggests that there exists significant constraint on the expansion of intron lengths among highly expressed zygotic genes during early fly development.

**DISCUSSION**

*Genome-wide Intron Delay in* Drosophila

The results of our analysis indicate that intron delay plays a significant role in determining patterns of expression in the early development of *Drosophila*: the production of long transcripts is limited by the rapid syncytial divisions occurring during zygotic genome activation. While a negative relationship between transcript length and expression level across a wide variety of organisms has been noted for some time [23], [24], this cannot explain our observation that in all three timecourses, the magnitude of the difference in expression level between the two length categories of genes declines across development. Furthermore, our observations are inconsistent with reduced transcriptional initiation limiting the transcription of long zygotic transcripts as evidenced by the lack of explanatory patterns in well-studied activating or repressive chromatin marks as well as the declining 5′:3′ ratio of coverage over the earliest embryonic stages among these genes. The larger proportion of reads being derived from 5′ ends is consistent with an inability to complete transcription of long genes, leading to an absence of 3′-derived reads (Figure 3A).

While intron delay clearly places an upper limit on the ability to express long zygotic genes, the inability to complete transcription cannot be the sole factor limiting their early expression because no zygotic transcripts are detected prior to syncytial cycle 4, irrespective of their length [28]. Furthermore, experimental forced arrest of embryos in non-mitotic portions of the cell cycle does not lead to full zygotic activation prior to syncytial cycle 10 [29]. Therefore, it would appear that the earliest steps of zygotic activation require the action of genes involved in pre- and post-translational processes, and are tightly linked to the programmed degradation of maternal RNAs [30]. While the intron delay hypothesis originally focused on the earliest periods of development, in both the RNA-Seq data (Figure 2A) and four-species microarray data (Figure 4) the expression of long zygotic genes continues to increase faster than short genes well into embryogenesis (~12-18 h). This may be explained by the observation that after gastrulation, large portions of the embryo form into mitotic domains [16] that begin amplifying their genomic content via endocycling – replication of all or parts of the genome via a modified cell cycle that bypasses mitosis as well as large portions of the gap phases to produce polyploid nuclei [17]. This modified cell cycle may be shortened, and therefore have the potential to physically limit long zygotic transcripts from achieving maximal expression until mid-embryogenesis. This hypothesis is also consistent with the sharp increase in 3′ derived reads observed among non-poly-A selected RNA-Seq data in the latter half of embryogenesis (Document S1, Figure S4).

*Embryonic expression across* Drosophila

At present we only have information on the maternally deposited transcriptome for *D. melanogaster*. However when maternal and zygotic gene classifications from *D. melanogaster* are applied to species up to 63 my diverged, patterns of embryonic expression remain qualitatively similar (Figure 4). The consistent, significant differences observed in expression patterns among short and long zygotic transcripts as well as maternal genes across the phylogeny suggest that the origin of these transcripts within the developing embryo may be largely conserved (see below).

As expected, early zygotic transcripts that are highly expressed across *Drosophila* are significantly shorter than those that are expressed at low levels or maternally deposited (Figure S6). It is interesting to note that while high early zygotic expression necessitates short primary transcripts, the converse does not hold: the range of intron lengths spanned by genes with low levels of expression (62 - 61,000 bp) is much greater than that spanned by highly expressed genes (52 – 3,600 bp). Nevertheless, our observation that the introns of highly expressed early zygotic genes have remained short across *Drosophila* species argues that the biological requirement of maintaining high levels of expression during early development is a major selective 'force' acting to maintain such conserved length. This is also supported by our observation that none of the other transcript categories – lowly expressed zygotic or either category of maternal transcripts – show significant differences in their variability across species (Figure 5).

*Maternal deposition vs. zygotic transcription*

Weischaus [31] hypothesized that "[i]n organisms where embryonic development is rapid and occurs with no increase in size before hatching from the egg, it will be advantageous to maximize maternal contributions, because the duration of oogenesis is often much longer than embryogenesis and the ovary provides a more sophisticated and efficient synthetic machinery." Despite the potential advantage of accelerated development, a significant fraction of the transcripts expressed in the embryos of species that fit the predictions of the above model appear to originate zygotically: 30-35% in *D. melanogaster* [22] and ~30% in the nematode *Caenorhabditis elegans* [32]. One explanation for the retention of zygotic origin is that a substantial fraction of these transcripts appears to require precise spatial localization [28] (especially if their unintended presence is deleterious to development [31]), which may limit their ability to transition to diffuse maternal deposition. In addition, short zygotically expressed genes may derive little or no benefit from being maternally supplied, as we observe that they are able to reach substantial levels of expression during early stages. Maternal deposition as proposed above may therefore only be advantageous in the case of long genes, which could bypass the expression constraints imposed upon them by intron delay. Supporting this possibility, genes expressed during the embryonic timecourse whose processed mRNAs are ≥ 5 kb are significantly over-represented among maternal as compared to zygotic genes (567 versus 239, $\chi^2 = 125.89$, 1 d.f., $p < 0.0001$). Determining which transcripts are supplied maternally versus expressed zygotically in sufficiently closely-related species would allow us to establish whether transitions to maternal deposition are common, and whether such events favor particular types of genes (e.g., those with long pre-processed transcripts).

## Conclusion

Intron delay appears to play a significant, yet underappreciated, role in determining patterns of expression beginning from the earliest moment of *Drosophila* embryogenesis, leading to clear expectations that zygotic mRNAs derived from long primary transcripts may take several hours after zygotic genome activation to reach full expression levels. This is an appealing mechanism through which to delay expression of transcripts that require precise temporal and spatial regulation of transcription until the necessary embryonic patterning gradients are established [1]. At present, it is difficult to rule out the possibility that delayed expression of long genes may also be under direct transcriptional control in addition to being subject to intron delay; however, in the case of at least one pair of *D. melanogaster* genes, *knirps* and *knrl*, the latter's delayed expression can be explained entirely by its long length [13]. As we continue to decipher the regulatory logic underlying transcription, we should be able to identify candidate genes whose long introns could be experimentally deleted and assessed for similar elimination of delay. Information gleaned from a sufficiently large sample of such genes will allow us to determine to what degree intron delay is used as an active mechanism of temporal regulation.

## MATERIALS & METHODS

RNA and ChIP-seq data

Mapped data from Gravely et al.'s [18] timecourse for each of the 12 time points spanning embryogenesis were obtained from the ModENCODE Data Coordination Center (http://www.modencode.org/; datasets modENCODE_2884 to modENCODE_2895). Counts of all 15,233 annotated loci (excluding pseudogenes and microRNA precursors) with FlyBase gene identification numbers (FBgns) in the FlyBase *D. melanogaster* genome annotation release 5.43 (FBr5.43) [27] were calculated using HTseq-count at the gene level with the 'union' option (http://www-huber.embl.de/users/anders/HTSeq/doc/index.html). Data were normalized by conditional quantile normalization using the 'cqn' Bioconductor package in R version 2.14 [33] and expression levels were output as RPKM. The raw RNA-Seq reads from (Lott et al. 2011) were obtained from the National Center for Biotechnology Information's (NCBI) Gene Expression Omnibus (GEO) (http://www.ncbi.nlm.nih.gov/geo/; accession GSE25180) and mapped to the FlyBase *D. melanogaster* genome release 5 using Tophat 1.0.13 [34] with default settings with the exception of a minimum intron length of 42 and retaining only uniquely mapping reads. Sexed data for each stage were collapsed and counting, normalization, and RPKM calculation were performed as for the Gravely et al. [18] dataset. In both datasets, we required that a gene be expressed at RPKM > 5 during at least one stage in order to be considered for analysis, leaving 10,454 loci in the Gravely et al. [18] dataset and 7,223 loci in the Lott et al. [19] dataset (lowering the threshold of expression to RPKM > 2 had no effect on our conclusions; data not shown).

We obtained the maternal and zygotic gene classifications of Tadros et al. [22] as tabulated in NCBI GEO entry GSE8910. All loci represented on the microarray platform used in the study (GPL1467) were converted to current FBgns using the FlyBase batch download tool. Those loci that were no longer part of the current annotation were excluded, while instances where multiple loci had been collapsed into a single locus in the current annotation were inspected to determine whether all collapsed loci were originally classified into the same category (i.e., maternal or zygotic). All cases where collapsed loci disagreed in terms of classification were rejected, providing 9,078 loci in the FBr5.43 annotation classified as maternally deposited or zygotically expressed, of which 7,452 (4,575 [61%] maternal/2,877 [39%] zygotic) were expressed in the Gravely et al. [18] dataset and 5,644 (4,151 [74%] maternal/1,493 [26%] zygotic) were expressed in the Lott et al. [19] dataset.

For the analysis of the distribution of reads on the 5′and 3′ ends of transcripts, we obtained the non-poly-A selected embryonic timecourse RNA-Seq reads generated by a SOLiD instrument (Life Technologies, Carlsbad, California) from the NCBI Short-read archive (SRA Accession numbers: SRX015641 to SRX015652) [18]. Reads were mapped to the *D. melanogaster* genome using the same methods as those applied to the syncytial timecourse of Lott et al. [19]. Using a custom script, combined with HTseq-count at the locus level with the ‘union’ option, we counted the number of reads spanning the 5′ and 3′ 1 kb of each transcript excluding any intronic sequence. Because non-poly-A selected RNA contains a mixture of both processed and unprocessed pre-mRNAs we chose to look only at those sequence segments that would be consistent between these two categories. As the segments analyzed were too short to perform conditional quantile normalization as above, read counts were quantile normalized using the R aroma.light package [35] and RPKMs calculated. We then calculated the 5′:3′ ratio for each transcript that a) was included as part of the embryonic timecourse (see criteria above), b) had a transcript of at least 2 kb in length, c) had only a single TSS according to the FlyBase 5.43 annotation and d) did not overlap another transcript leaving 3,396 loci with 5′:3′ ratios to analyze.

Raw developmental timecourse ChIP-seq reads derived from antibodies to histone modifications H3K4me3, H3K9Ac, H3K27me3, and H3K9me3 [25] were obtained from GEO (accession numbers to all datasets are found in Table S2). All reads were mapped uniquely to the FlyBase *D. melanogaster* genome release 5 using Bowtie version 0.12.8 and allowing 2 mismatches [36]. Base-level coverage was assessed in 100 bp non-overlapping windows up to one kb upstream of non-overlapping genes with a single annotated TSS. Coverage was normalized between time points and chromatin marks by dividing by the total number of mapped reads by $10^6$.

Choice of ‘short’ and ‘long’ locus categories

In order to determine appropriate transcript length cutoffs to detect the potential effect of intron delay, we first began by binning all loci in the FlyBase 5.43 annotation into increments of 5 kb (i.e., 5, 10, 15, 20 kb, etc.). Visual inspection of the pattern of expression of the length categories indicated an increasing degree of effect (i.e., progressively longer bins showed a more pronounced reduction in expression during early vs. later stages of development; data not shown). We then performed pairwise comparisons of the distributions of expression levels of the individual bins during each of the time points and found that there were no significant differences among those bins with loci > 5 kb in length ($p > 0.05$) whereas these same bins were significantly different from those loci < 5 kb in length. Therefore we defined two length categories, short (< 5kb) and long (≥ 5 kb), whose expression patterns were significantly different from one another.

GO analysis

All maternal or zygotic loci considered significantly expressed in either the embryonic or syncytial timecourses were analyzed for functional over- and under-representation using FatiGO [37] on the Babelomics version 4.3 webserver at (http://babelomics.bioinfo.cipf.es/functional.html). Gene lists were compared either to one another or the whole FlyBase 5.43 annotation among GO biological process levels from three to nine using two-tailed tests and retaining only p values < 0.05 when adjusted for multiple tests by the software.

Four-species microarray data

We obtained the processed microarray data as described in Kalinka et al. [26] from http://publications.mpi-cbg.de/getDocument.html?id=ff8080812c477bb6012c5fa1feaf0047. All locus names were associated with *D. melanogaster* FlyBase FBgns. Loci from the Kalinka et al. dataset, which was based on FlyBase annotation 5.14, that were not associated with unique FBgns (either due to a locus having been split into multiple loci or multiple loci having collapsed into a single locus in the FBr5.43 annotation used in this study) were removed from further analysis. As two species, *D. simulans* and *D. persimilis*, were originally noted to have poor genome sequencing coverage [38], we used the remaining FBgns to search for orthologs in *D. ananassae* (FlyBase genome release 1.3, annotation release FB2011_07), *D. pseudoobscura* (FlyBase genome release 2.27, annotation release FB2012_02), and *D. virilis* (FlyBase genome release 1.2, annotation release FB2012_01) using the FlyBase batch download tool. Of the 3,146 loci mapping to a single ortholog in all three *non-melanogaster* species, 2,067 were represented among the *D. melanogaster* zygotic and maternal loci annotated by Tadros et al. [22]. These loci were retained for further analysis and were called maternal or zygotic based on the *D. melanogaster* data. We used the average normalized, processed expression level among all probes represented over time points 1-9 for each locus within a given species for analysis as data was not available for all species for any subsequent time points.

Orthologous intron analysis

As the genome annotations of non-*melanogaster* species of *Drosophila* largely lack untranslated regions as well as alternatively spliced isoforms that could lead to changes in primary transcript length, we sought to compare only orthologous intronic segments. These segments were identified using the software Common Introns Within Orthologous Genes (CIWOG) [39] on the genome releases indicated above retaining only those segments that were common among all four species analyzed. In order to compare variability in intron lengths, we used the corrected coefficient of variation (CV*) [40], removing all single exon genes. It should be noted that CV* is biased towards low values when mean intron lengths among species are < 150 bp (data not shown) as is the case with most introns in the highly expressed zygotic category (Figure 5). Upon reanalyzing the data after removing all loci with mean intron length among species < 150 bp, the only significant difference in CV* is observed among highly expressed zygotic and low expressed maternal transcripts ($p < 0.01$). However, this serves to indicate that the transcript length range tolerated by highly expressed zygotic loci during early development is short and narrow relative to other locus categories.

General statistics

All statistics were performed using R version 2.14.0 [41]. Confidence intervals were obtained by producing a normal approximation of 10,000 resampled subsets of the data using the ‘boot’ package in R [42]. Comparisons between distributions were performed using the permuted Kruskal-Wallis rank sum test, with 10,000 permutations, as implemented in the ‘coin’ package in R [43]. The p-values of all comparisons were Bonferroni corrected for multiple tests where appropriate

## ACKNOWLEDGEMENTS

We thank T. Babak, J. Walters, A. Bergland, as well as members of the Fraser and Petrov labs for useful comments on earlier versions of this manuscript.

## REFERENCES

1. Gubb D. (1986) Intron-delay and the precision of expression of homoeotic gene products in Drosophila. Dev Genet 7: 119–131.

2. Swinburne IA, Silver PA. (2008) Intron Delays and Transcriptional Timing during Development. Dev Cell 14: 324-330.

3. Tennyson CN, Klamut HJ, Worton RG. (1995) The human dystrophin gene requires 16 hours to be transcribed and is cotranscriptionally spliced. Nat Genet 9: 184–190.

4. Jeffares DC, Penkett CJ, Bähler J. (2008) Rapidly regulated genes are intron poor. Trends Genet 24: 375–378.

5. Gottesfeld JM, Forbes DJ. (1997) Mitotic repression of the transcriptional machinery. Trends Biochem Sci 22: 197–202.

6. Shermoen AW, O'Farrell PH. (1991) Progression of the cell cycle through mitosis leads to abortion of nascent transcripts. Cell 67: 303–310.

7. Mourier T, Jeffares DC. (2003) Eukaryotic intron loss. Science 300: 1393.

8. Jeffares DC, Mourier T, Penny D. (2006) The biology of intron gain and loss. Trends Genet 22: 16–22.

9. Foe VE, Odell GM, Edgar BA. (1993) Mitosis and morphogenesis in the Drosophila embryo: Point and counterpoint. In: Bate M, Martinez Arias A, editors. The Development of *Drosophila melanogaster*. Cold Spring Harbor: Cold Spring Harbor Laboratory Press. pp. 149-300.

10. Grbić M, Nagy LM, Carroll SB, Strand M. (1996) Polyembryonic development: insect pattern formation in a cellularized environment. Development 122: 795–804.

11. Foe VE, Alberts BM. (1983) Studies of nuclear and cytoplasmic behaviour during the five mitotic cycles that precede gastrulation in Drosophila embryogenesis. J Cell Sci 61: 31–70.

12. Frescas D, Mavrakis M, Lorenz H, Delotto R, Lippincott-Schwartz J. (2006) The secretory membrane system in the Drosophila syncytial blastoderm embryo exists as functionally compartmentalized units around individual nuclei. J Cell Biol 173: 219–230.

13. Rothe M, Pehl M, Taubert H, Jäckle H. (1992) Loss of gene function through rapid mitotic cycles in the Drosophila embryo. Nature 359: 156–159.

14. De Renzis S, Elemento O, Tavazoie S, Wieschaus EF. (2007) Unmasking activation of the zygotic genome using chromosomal deletions in the Drosophila embryo. PLoS Biol 5: e117.

15. Gubb D. (1998) Cellular polarity, mitotic synchrony and axes of symmetry during growth. Where does the information come from? Int J Dev Biol 42: 369–377.

16. Foe VE. (1989) Mitotic domains reveal early commitment of cells in Drosophila embryos. Development 107: 1–22.

17. Edgar BA, Orr-Weaver TL. (2001) Endoreplication cell cycles: more for less. Cell 105: 297–306.

18. Graveley BR, Brooks AN, Carlson JW, Duff MO, Landolin JM, et al. (2011) The developmental transcriptome of Drosophila melanogaster. Nature 471: 473–479.

19. Lott SE, Villalta JE, Schroth GP, Luo S, Tonkin LA, et al. (2011) Noncanonical compensation of zygotic X transcription in early Drosophila melanogaster development revealed through single-embryo RNA-seq. PLoS Biol 9: e1000590.

20. Campos-Ortega JA, Hartenstein V. (1985) The embryonic development of Drosophila melanogaster. Berlin: Springer-Verlag 405 p.

21. Walser CB, Lipshitz HD. (2011) Transcript clearance during the maternal-to-zygotic transition. Curr Opin Genetics Dev 21: 431–443.

22. Tadros W, Goldman AL, Babak T, Menzies F, Vardy L, et al. (2007) SMAUG is a major regulator of maternal mRNA destabilization in Drosophila and its translation is activated by the PAN GU kinase. Dev Cell 12: 143–155.

23. Duret L, Mouchiroud D. (1999) Expression pattern and, surprisingly, gene length shape codon usage in Caenorhabditis, Drosophila, and Arabidopsis. Proc Natl Acad Sci USA 96: 4482–4487.

24. Castillo-Davis CI, Mekhedov SL, Hartl DL, Koonin EV, Kondrashov FA. 2002. Selection for short introns in highly expressed genes. Nat Genet 31: 415–418.

25. modENCODE Consortium, Roy S, Ernst J, Kharchenko PV, Kheradpour P, et al. (2010) Identification of functional elements and regulatory circuits by Drosophila modENCODE. Science 330: 1787–1797.

26. Kalinka AT, Varga KM, Gerrard DT, Preibisch S, Corcoran DL, et al. (2010) Gene expression divergence recapitulates the developmental hourglass model. Nature 468: 811–814.

27. McQuilton P, St Pierre SE, Thurmond J, FlyBase Consortium. (2012) FlyBase 101--the basics of navigating FlyBase. Nucleic Acids Res 40: D706–714.

28. Lécuyer E, Yoshida H, Parthasarathy N, Alm C, Babak T, et al. (2007) Global analysis of mRNA localization reveals a prominent role in organizing cellular architecture and function. Cell 131: 174–187.

29. Edgar BA, Schubiger G. (1986) Parameters controlling transcriptional activation during early Drosophila development. Cell 44: 871–877.

30. Tadros W, Lipshitz HD. (2009) The maternal-to-zygotic transition: a play in two acts. Development 136: 3033–3042.

31. Wieschaus E. (1996) Embryonic Transcription and the Control of Developmental Pathways. Genetics 142: 5.

32. Baugh LR, Hill AA, Slonim DK, Brown EL, Hunter CP. (2003) Composition and dynamics of the Caenorhabditis elegans early embryonic transcriptome. Development 130: 889–900.

33. Hansen KD, Irizarry RA, Wu Z. 2012. Removing technical variability in RNA-seq data using conditional quantile normalization. Biostatistics 13: 204–216.

34. Trapnell C, Pachter L, Salzberg SL. (2009) TopHat: discovering splice junctions with RNA-Seq. Bioinformatics 25: 1105–1111.

35. Bengtsson H, Hössjer O. (2006) Methodological study of affine transformations of gene expression data with proposed robust non-parametric multi-dimensional normalization method. BMC Bioinformatics 7: 100.

36. Langmead B, Trapnell C, Pop M, Salzberg SL. (2009) Ultrafast and memory-efficient alignment of short DNA sequences to the human genome. Genome Biol 10: R25.

37. Al-Shahrour F, Minguez P, Tárraga J, Medina I, Alloza E, et al. (2007) FatiGO +: a functional profiling tool for genomic data. Integration of functional annotation, regulatory motifs and interaction data with microarray experiments. Nucleic Acids Res 35: W91–96.

38. Drosophila 12 Genomes Consortium, Clark AG, Eisen MB, Smith DR, Bergman CM, et al. (2007) Evolution of genes and genomes on the Drosophila phylogeny. Nature 450: 203–218.

39. Wilkerson MD, Ru Y, Brendel VP. (2009) Common introns within orthologous genes: software and application to plants. Brief Bioinformatics 10: 631–644.

40. Sokal RR, Rohlf FJ. (1995) Biometry: The Principles and Practice of Statistics in Biological Research, 3rd ed. New York: W. H. Freeman and Company. 880 p.

41. R Development Core Team. (2008) R: A language and environment for statistical computing. Vienna: R Foundation for Statistical Computing. URL http://www.R-project.org.

42. Davison AC, Hinkley DV. (1997) Bootstrap Methods and Their Applications. Cambridge: Cambridge University Press. 594 p.

43. Hothorn T, Hornik K, van de Wiel MA, Zeileis A. (2008) Implementing a Class of Permutation Tests: The coin Package. J Stat Software 28: 1-23.

44. Thomsen S, Anders S, Janga SC, Huber W, Alonso CR. (2010) Genome-wide analysis of mRNA decay patterns during early Drosophila development. Genome Biol 11: R93.

45. Beyer AL, Osheim YN. (1988) Splice site selection, rate of splicing, and alternative splicing on nascent transcripts. Genes Dev 2: 754–765.

46. Osheim Y, and Beyer A. (1989) Electron Microscopy of Ribonucleoprotein Complexes on Nascent RNA using Miller Chromatin Spreading Method. Method enzymol 180: 481–509.

47. Yin H, Sweeney S, Raha D, Snyder M, Lin H. (2011) A high-resolution whole-genome map of key chromatin modifications in the adult Drosophila melanogaster. PLoS Genet 7: e1002380.

48. Vakoc CR, Mandat SA, Olenchock BA, Blobel GA. (2005) Histone H3 lysine 9 methylation and HP1gamma are associated with transcription elongation through mammalian chromatin. Mol Cell 19: 381–391.

## FIGURE LEGENDS

**Figure 1**. Correspondence between the three expression timecourses analyzed in this study: Embryonic [18], species [26], and syncytial [19]. Both the embryonic and species timecourses consist of pools of embryos collected at two hour intervals, spanning either 24 or 18 h of *Drosophila* embryogenesis. The syncytial timecourse spans syncytial cycles 10 to 13, followed by 4 collections during the extended 14th cycle corresponding roughly to 25% increments of cell wall extension to completion of cellularization (indicated by A-D). The correspondence between the syncytial timecourse and the other timecourses is indicated by the grey dotted line. The hashed area indicates the period during which the rapid syncytial divisions take place (timing is taken from [20]). The embryonic and syncytial timecourses were generated by RNA-Seq while the species timecourse was generated using microarrays.

**Figure 2**. Median expression levels (with 95% confidence intervals) for zygotic (black) and maternal (red) genes over the embryonic (A and B) and syncytial (C and D) timecourses, respectively. Short (< 5 kb) and long genes (≥ 5 kb) are indicated as circles and triangles, respectively. Median expression levels of both zygotic gene length classes increase over both the embryonic and syncytial timecourses, however the difference in expression level between two length categories becomes smaller over subsequent stages of development, as predicted by the intron delay hypothesis. Neither length category increases significantly among maternal genes.

**Figure 3.** (A) Illustration of the predicted read coverage (indicated as grey bars) along short and long transcripts under *cis*-regulation vs. intron delay models explaining the lower expression of long zygotic transcript during early development. Under a regulatory model, the 5′ and 3′ ends of all transcripts should have relatively similar read coverage. Under the intron delay model, however, the 5′:3′ ratio of long genes should be > 1 during early development, and decrease as development progresses. Median 5′:3′ ratios over the embryonic timecourse as determined from total RNA SOLiD data are indicated for zygotic (B) and maternal (C) genes in black and red, respectively. Short (< 5 kb) and long (≥ 5 kb) genes are indicated as circles and triangles, respectively. The 5′:3′ ratio shows a decrease over the first six hours of development in long zygotic genes, but not in any other category, as expected under the predictions of the intron delay model.

**Figure 4.** Median expression levels (with 95% confidence intervals) for zygotic (black) and maternal (red) genes among the species of the four species microarray-based timecourse. Short genes (< 5 kb) and long genes (≥ 5 kb) are indicated as circles and triangles, respectively. All three non-*melanogaster* species show patterns consistent with the RNA-Seq based embryonic timecourse.

**Figure 5.** Corrected coefficients of variation (CV*) in orthologous intron length among the four species analyzed for high and low expression genes during the 0-2 h time point of the species timecourse among zygotic (black) and maternal (red) genes. The only significant difference among distributions is the comparison between the high expression zygotic gene category and all others.

## TABLES

**Table 1.** Summary statistics for the embryonic, syncytial, and species timecourses. Median primary transcript length is shown with bootstrapped 95% confidence intervals. Note that the species timecourse classifications and median length were calculated using mean orthologous intron lengths across the four species analyzed.

| Timecourse | Expression Origin | Length Category | Number of Genes | Median Primary Transcript Length (bp) |
|---|---|---|---|---|
| Embryonic | Zygotic | Short (< 5 kb) | 2,100 | 1,699 ± 61 |
| | | Long (≥ 5kb) | 777 | 11,300 ± 1,121 |
| | Maternal | Short (< 5 kb) | 3,058 | 2222 ± 64 |
| | | Long (≥ 5kb) | 1,517 | 10,303 ± 485 |
| Syncytial | Zygotic | Short (< 5 kb) | 1,184 | 1,690 ± 78 |
| | | Long (≥ 5kb) | 309 | 10,365 ± 1,615 |
| | Maternal | Short (< 5 kb) | 2,843 | 2,242 ± 66 |
| | | Long (≥ 5kb) | 1,308 | 9,702 ± 587 |
| Four Species | Zygotic | Short (< 5 kb) | 684 | 191 ± 28 |
| | | Long (≥ 5kb) | 93 | 10,933 ± 2,092 |
| | Maternal | Short (< 5 kb) | 1,142 | 242 ± 20 |
| | | Long (≥ 5kb) | 148 | 10,244 ± 2001 |

Figure 1

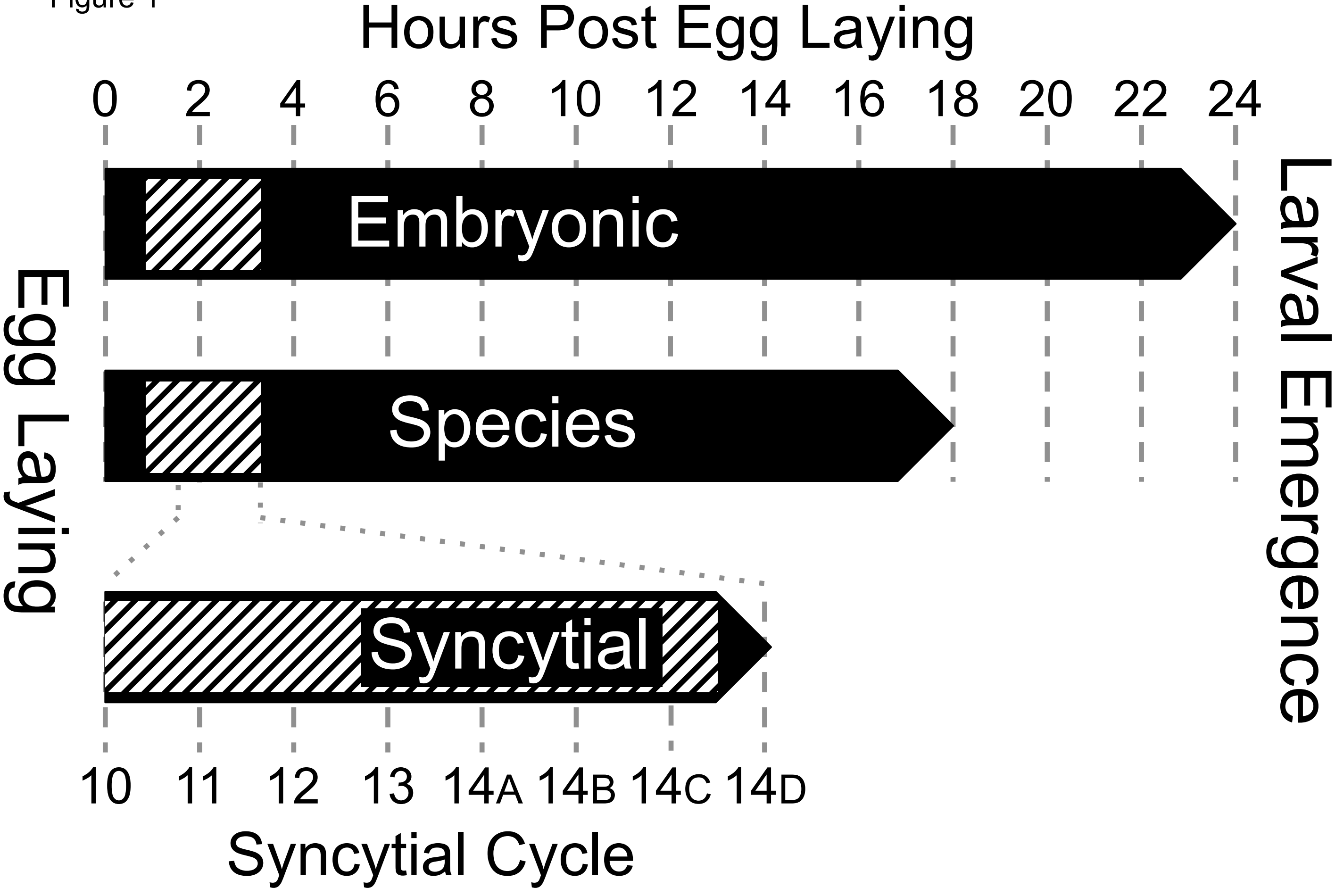

Figure 2

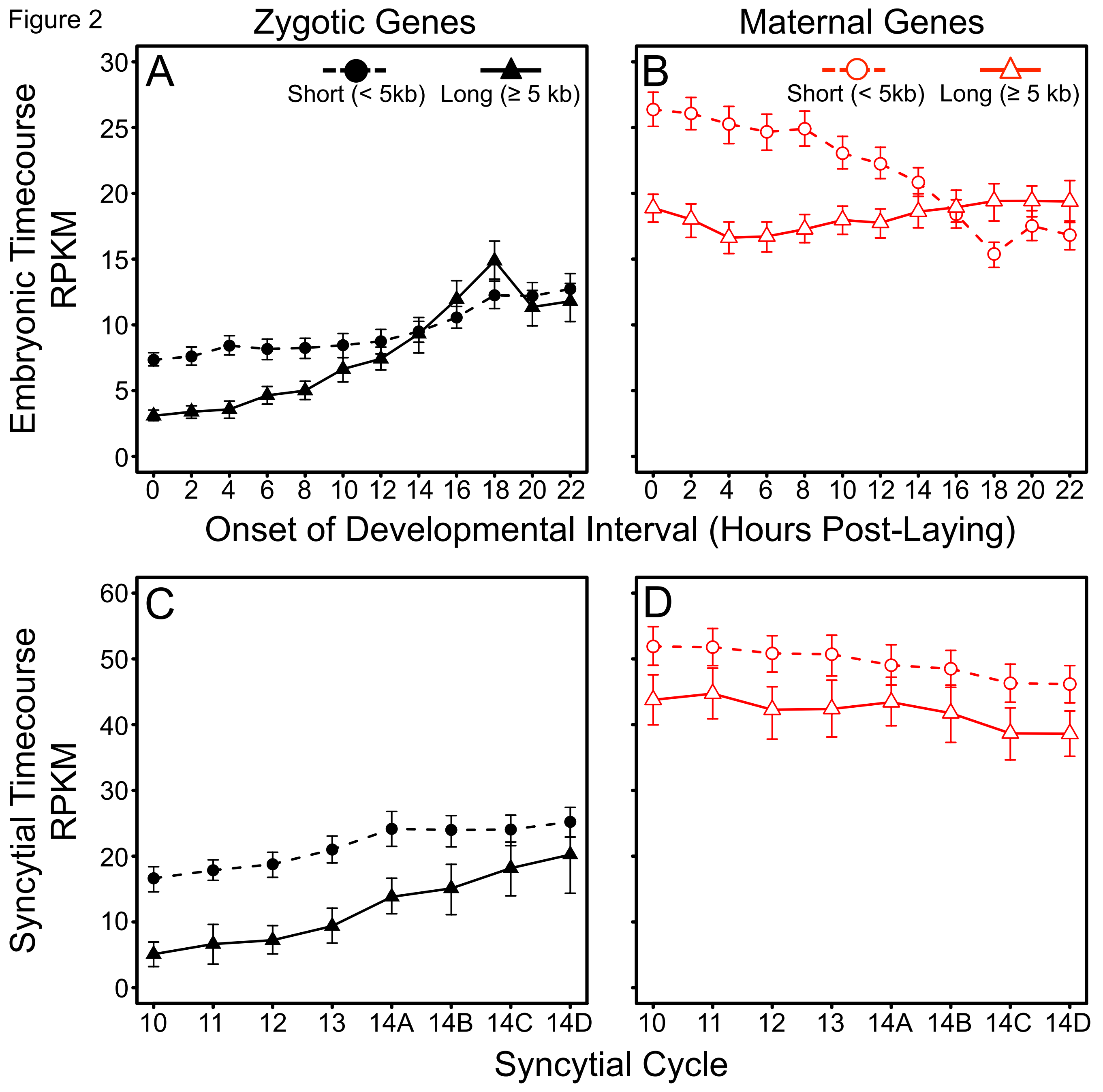

Figure 3

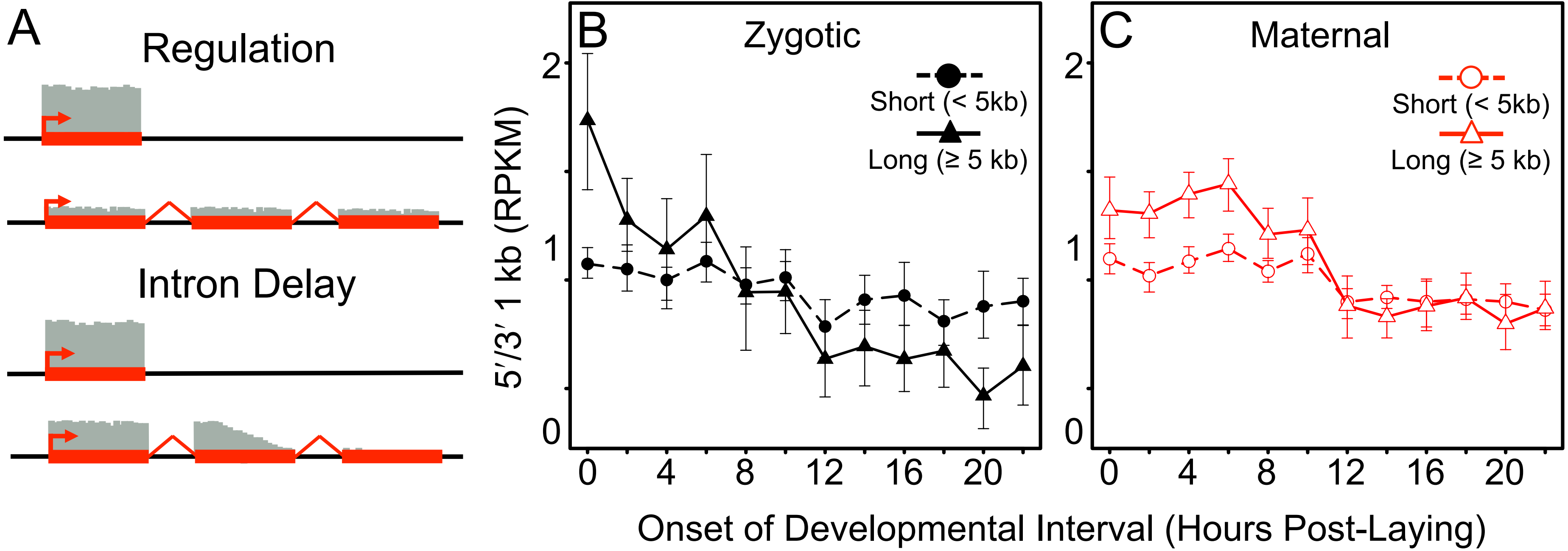

Figure 4

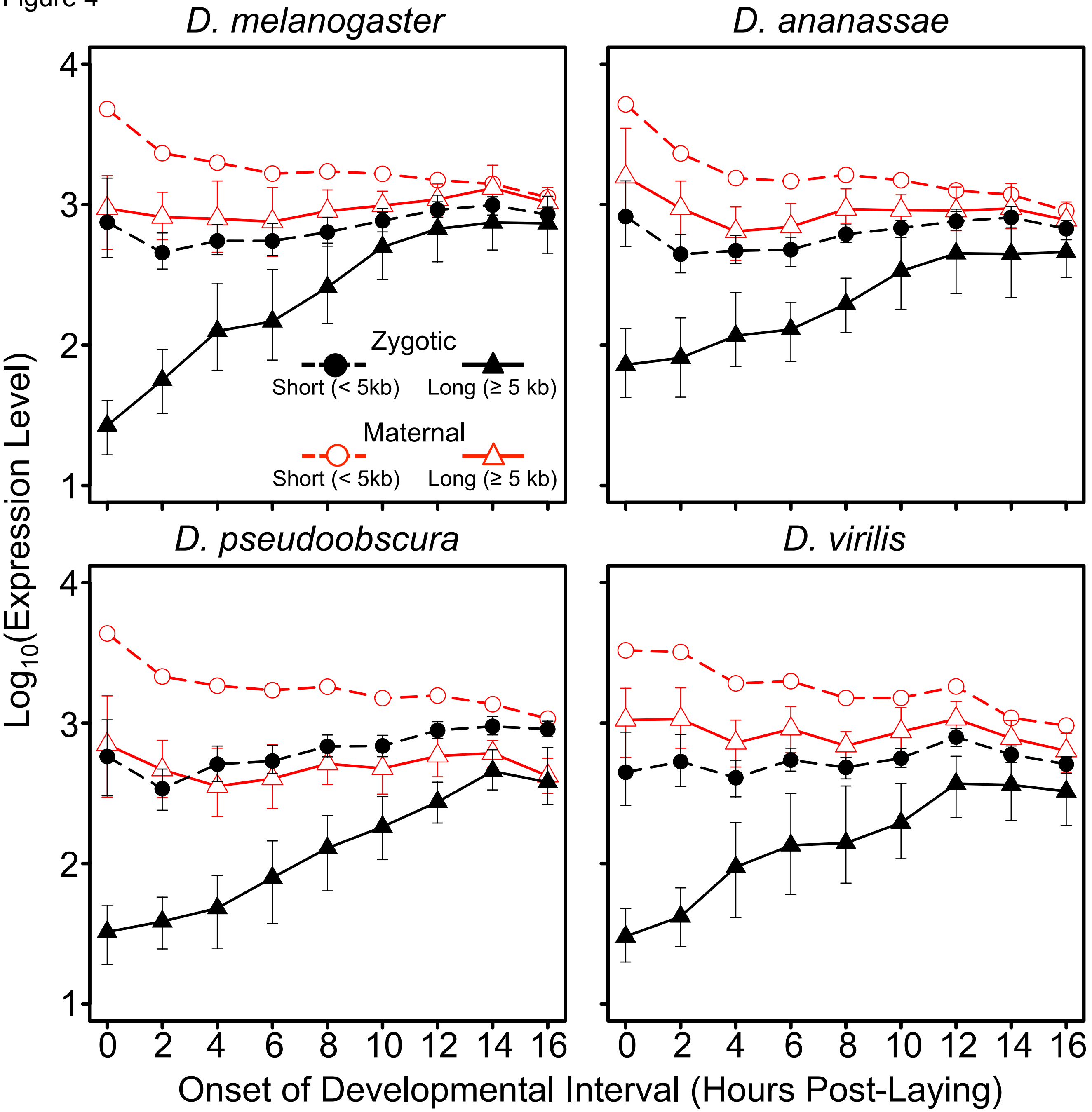

Figure 5

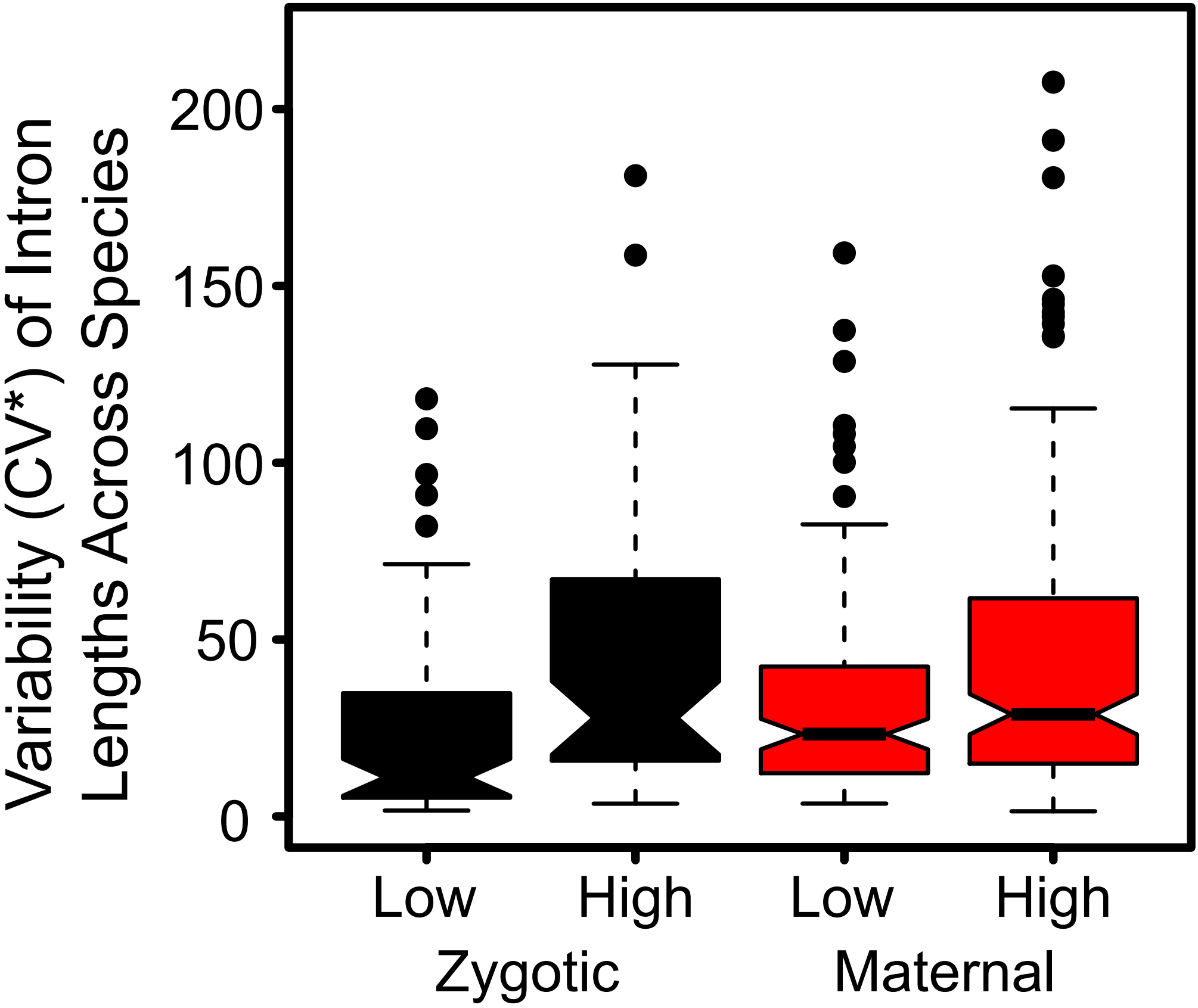

**SUPPLEMENTARY INFORMATION**

This document contains supplementary results, analysis, figures, and tables not included in the main manuscript.

*Transcript length and zygotic origin are not associated with particular functional classes*

In order to better characterize potential functional differences among transcripts, we analyzed the representation of the different size categories of maternal and zygotic genes among Gene Ontology (GO) biological process terms using FatiGO [36]. Consistent with previous analyses [22], [43], maternally deposited genes are over-represented as compared to the genome as a whole in a number of functional classes (Table S3). In contrast, we found that zygotic genes did not show significant over-representation among any GO biological process categories ($p > 0.05$). Furthermore, when zygotic and maternal genes were compared to one another, neither long nor short transcripts were significantly over-represented in any GO category. Consequently, there is not a clear set of functions associated with whether zygotic genes are subject to or escape intron delay.

*Rejection of a kinetic mechanism restricting early expression of long zygotic transcripts*

A third potential explanation for the low expression level of long transcripts followed by a progressive reduction in the relative ratio of median expression of short/long genes is transcriptional kinetics. Consider a simple model under which a single RNA Pol II complex transcribes a locus at any one time. Assuming that expression level eventually reaches saturation, multiple short transcripts could be produced in the same amount of time required to transcribe a long transcript, allowing short genes to reach saturation while delaying maximal expression of longer genes. However, there are several lines of evidence that argue against this possibility. First, while it is not known whether co-transcription of multiple RNA Pol II complexes is a common feature of long genes in general, it has been observed *in situ* at a number of specific loci, suggesting that such a simple model does not capture biological reality (e.g., [45], [46]). Second, a purely kinetic model cannot explain the decreasing 5′:3′ ratio observed among long zygotic transcripts (Figure 3B) without the addition of degradation of incomplete transcripts upon mitosis as required by the intron delay model. Finally, if expression level was primarily explained by kinetics, we would expect to observe a strong negative relationship between length and transcript abundance *within* developmental stages of the embryonic timecourse, and most especially during the earliest stages. While the strongest negative relationship between zygotic gene length and expression level does occur during the first embryonic stage (0-2 h) (Figure S1), the proportion of the variance explained by locus length is low ($R^2 = 0.0658$). Thus while it impossible to exclude the possibility that transcriptional kinetics are playing some role in the time required for long loci to reach stable levels of expression, it cannot be the primary determinant of long locus expression delay.

*Lack of evidence for differences in the abundance of chromatin marks associated with activation or repression in short vs. long zygotic genes*

As an alternative to the intron delay hypothesis, long zygotic genes could be preferentially delayed in activation only during early development by purely transcriptional means. This could manifest itself in chromatin structure profiles [46] via two non-mutually exclusive mechanisms: 1) Long zygotic genes could show a paucity of chromatin marks indicating active transcription (such as H3K4me3 and H3K9Ac) relative to short zygotic genes during early development, with such marks increasing over time reflecting increased expression, and 2) Long zygotic genes could show an excess of repressive or heterochromatic chromatin marks (such as H3K27me3 and H3K9me3) relative to short zygotic genes during early development, with this excess decreasing over time. In order to test this potential explanation of our observations, we obtained ChIP-Seq data from the modENCODE Consortium [25] generated using embryos collected in six four-hour windows over embryogenesis for the four histone H3 chromatin marks indicated above. Their relative coverage was determined in non-overlapping 100 bp windows for the 1 kb upstream of TSSs in our dataset (see Methods).

We did not observe a monotonic increase in the presence of euchromatic marks (H3K4me3 and H3K9Ac) in the upstream regions of long relative to short zygotic genes, indicating that they cannot explain the delayed expression of the former (Figure S2). In the case of the heterochromatic marks, there is no evidence of increased abundance of H3K27me3 upstream of the TSSs of long zygotic genes. However, long zygotic genes do show increased abundance of H3K9me3 ChIP-Seq coverage in the window between 100 to 200 bp upstream of their TSS during the 0-4 h developmental stage, which disappears in subsequent stages (Kruskal-Wallis rank sum test, $p = 0.03$) (Figure S3). However it should be noted that while generally considered a mark of heterochromatin, H3K9me3 has also been associated with active transcription [48]. Regardless, we tested the possibility that long zygotic genes were delayed in activation due to active repression early in development by repeating the analysis shown in Figure 2, excluding the long zygotic genes with an above-median level of normalized H3K9me3 coverage 100-200 bp upstream of the TSS. Our results remain qualitatively unchanged, and long zygotic genes continue to show significantly delayed activation as determined by ANCOVA ($F_{3,8}$, $p = 0.012$). Consequently, repression of expression as evidenced by increased H3K9me3 abundance during early development cannot explain our observations.

*Analysis of 5′:3′ ratios over embryogenesis*

Median 5′:3′ ratios were generally higher during the 0-12 h time points than in >12-24 h (Kruskal-Wallis rank sum test, $p < 2.2 \times 10^{-16}$ in all cases). Further analysis of this difference (Figure S4) revealed that long genes show a significantly higher 3′ RPKM during the latter half of embryogenesis, at which point they are being supplied entirely zygotically, contributing to the large change in 5′:3′ ratios observed in Figure 3 (see Discussion in main manuscript).

In order to confirm that the patterns observed in Figure 3 were not an artifact of examining only the terminal ends of transcripts, we used the same dataset to plot median base-level exonic coverage (normalized as a fraction of maximum coverage) in non-overlapping 500 bp windows over the 5′ most 10 kb of zygotic and maternal transcripts (Figure S5). Some caveats should be noted in this analysis: The majority of *D. melanogaster* transcripts in the dataset are < 5kb (~70%) (Table 1) leading to a substantially reduced amount of data for more distal windows. Compounding this, the majority of the sequence of long transcripts is intronic, leading to a general decrease in likelihood that any window will contain exonic bases required for analysis with increasing distance from the TSS. Finally, the lack of poly-A selection applied to the libraries used for this analysis vastly increases the number of RNA-seq reads derived from rRNA relative to mRNA, especially as compared to the primary datasets used in the embryonic and syncytial timecourses (Figure 2). Each of these factors contributes noise to the estimated expression levels of each window, particularly in those windows >= 5 kb from the TSS. Nevertheless, the overall patterns observed are consistent with a greater paucity of 3′ reads among zygotic as compared to maternal transcripts only during the earliest time points of embryonic development. Furthermore, we observe no time points during which maternal transcripts show a greater paucity of 3′ coverage (Figure S5).

*Zygotic genes highly expressed in early development are short*

We calculated the mean orthologous intron length across the species analyzed in the four species timecourse, ignoring any single-exon genes, and compared short and long maternal and zygotic transcripts (Figure S6). As expected, both zygotic and maternal high expression genes have shorter mean intron lengths than their corresponding low expression genes ($p < 2.2 \times 10^{-16}$). Furthermore, the high expression zygotic genes have the shortest mean intron lengths overall ($p < 2.2 \times 10^{-16}$).

**Table S2.** GEO datasets used in the analysis of ChIP-Seq chromatin marks.

| | Euchromatic | | Heterochromatic | |
|---|---|---|---|---|
| **Time point** | **H3K4me3** | **H3K9Ac** | **H3K27me3** | **H3K9me3** |
| 0-4 h | GSM400657 | GSM401408 | GSM439448 | GSM430457 |
| 4-8 h | GSM400674 | GSM401405 | GSM439447 | GSM436456 |
| 8-12 h | GSM439446 | GSM432592 | GSM439446 | GSM439455 |
| 12-16 h | GSM432580 | GSM439458 | GSM439445 | GSM439454 |
| 16-20 h | GSM400658 | GSM401402 | GSM439444 | GSM439453 |
| 20-24 h | GSM400672 | GSM401424 | GSM439443 | GSM439452 |

**Table S3.** Complete list of all GO Biological Process terms significantly over-represented among maternal loci in comparison to the genome as a whole. The loci determined to be expressed during the embryonic and syncytial timecourses were analyzed separately. p-values are adjusted to reflect a false-discovery rate of 0.05.

**Embryonic Timecourse**

| GO Term ID | GO Term Name | Loci in Dataset | Percent Among Maternal Loci | Percent Among Entire Genome | Adjusted p-value |
|---|---|---|---|---|---|
| GO:0007049 | Cell Cycle | 642 | 3.96 | 3.03 | 4.43E-02 |
| GO:0006508 | Proteolysis | 920 | 5.79 | 4.3 | 1.83E-03 |
| GO:0006950 | Response to stress | 1009 | 6.34 | 4.72 | 1.09E-03 |
| GO:0009056 | Catabolic process | 1244 | 7.91 | 5.79 | 1.78E-04 |
| GO:0009266 | Response to temperature stimulus | 596 | 3.83 | 2.76 | 9.17E-03 |
| GO:0009409 | Response to cold | 550 | 3.61 | 2.53 | 5.40E-03 |
| GO:0009628 | Response to Abiotic stimulus | 710 | 4.48 | 3.32 | 8.17E-03 |
| GO:0030163 | Protein catabolic process | 936 | 5.95 | 4.36 | 9.31E-04 |
| GO:0042309 | homoiothermy | 548 | 3.61 | 2.51 | 5.02E-03 |
| GO:0042592 | Homeostatic process | 695 | 4.59 | 3.18 | 8.69E-04 |
| GO:0050826 | Response to freezing | 548 | 3.61 | 2.51 | 5.02E-03 |
| GO:0016070 | RNA metabolic process | 1305 | 8.13 | 6.12 | 3.71E-04 |
| GO:0007166 | cell surface receptor signaling pathway | 929 | 5.68 | 4.39 | 9.90E-03 |
| GO:0006350 | Transcription | 1086 | 6.8 | 5.09 | 9.11E-04 |
| GO:0006351 | transcription, DNA-dependent | 956 | 5.9 | 4.5 | 5.40E-03 |
| GO:0006355 | regulation of transcription, DNA-dependent | 900 | 5.68 | 4.2 | 1.83E-03 |
| GO:0019222 | regulation of metabolic process | 1362 | 8.55 | 6.37 | 1.78E-04 |
| GO:0045449 | Regulation of transcription | 995 | 6.32 | 4.63 | 7.98E-04 |
| GO:0006412 | Translation | 489 | 3.1 | 2.28 | 4.23E-02 |
| GO:0005975 | carbohydrate metabolic process | 526 | 3.3 | 2.46 | 4.73E-02 |
| GO:0016192 | vesicle-mediated transport | 470 | 3.04 | 2.17 | 2.49E-02 |
| GO:0007242 | intracellular signal transduction | 554 | 3.61 | 2.55 | 7.16E-03 |
| GO:0007399 | nervous system development | 672 | 4.28 | 3.12 | 6.92E-03 |
| GO:0009653 | anatomical structure morphogenesis | 1159 | 7.23 | 5.44 | 7.98E-04 |
| GO:0009887 | organ morphogenesis | 610 | 3.83 | 2.86 | 2.61E-02 |
| GO:0022008 | neurogenesis | 500 | 3.19 | 2.32 | 3.07E-02 |
| GO:0030154 | cell differentiation | 1072 | 6.86 | 4.98 | 2.62E-04 |
| GO:0009888 | tissue development | 563 | 3.54 | 2.63 | 3.25E-02 |
| GO:0006836 | neurotransmitter transport | 147 | 1.14 | 0.62 | 2.08E-02 |

## Syncytial Timecourse

| GO Term ID | GO Term Name | Loci in Dataset | Percent Among Maternal Loci | Percent Among Entire Genome | Adjusted p-value |
|---|---|---|---|---|---|
| GO:0007049 | Cell Cycle | 635 | 4.19 | 3.03 | 7.68E-03 |
| GO:0006508 | Proteolysis | 891 | 5.69 | 4.3 | 6.41E-03 |
| GO:0006950 | Response to stress | 982 | 6.34 | 4.72 | 2.22E-03 |
| GO:0009056 | Catabolic process | 1209 | 7.88 | 5.79 | 2.87E-04 |
| GO:0009266 | Response to temperature stimulus | 582 | 3.88 | 2.76 | 7.72E-03 |
| GO:0009409 | Response to cold | 538 | 3.69 | 2.53 | 4.23E-03 |
| GO:0009628 | Response to Abiotic stimulus | 694 | 4.55 | 3.32 | 6.41E-03 |
| GO:0030163 | Protein catabolic process | 905 | 5.81 | 4.36 | 5.15E-03 |
| GO:0042309 | homoiothermy | 536 | 3.69 | 2.51 | 3.74E-03 |
| GO:0042592 | Homeostatic process | 680 | 4.7 | 3.18 | 6.32E-04 |
| GO:0050826 | Response to freezing | 536 | 3.69 | 2.51 | 3.74E-03 |
| GO:0006396 | RNA processing | 289 | 2.02 | 1.35 | 3.46E-02 |
| GO:0016070 | RNA metabolic process | 1274 | 8.21 | 6.12 | 3.57E-04 |
| GO:0007166 | cell surface receptor signaling pathway | 904 | 5.66 | 4.39 | 1.71E-02 |
| GO:0006350 | Transcription | 1059 | 6.84 | 5.09 | 1.35E-03 |
| GO:0006351 | transcription, DNA-dependent | 933 | 5.95 | 4.5 | 5.53E-03 |
| GO:0006355 | regulation of transcription, DNA-dependent | 879 | 5.76 | 4.2 | 1.97E-03 |
| GO:0019222 | regulation of metabolic process | 1330 | 8.65 | 6.37 | 2.87E-04 |
| GO:0045449 | Regulation of transcription | 971 | 6.38 | 4.63 | 7.69E-04 |
| GO:0042221 | response to chemical stimulus | 362 | 2.43 | 1.71 | 4.39E-02 |
| GO:0005975 | carbohydrate metabolic process | 515 | 3.37 | 2.46 | 3.23E-02 |
| GO:0006812 | cation transport | 394 | 2.65 | 1.86 | 3.46E-02 |
| GO:0016044 | cellular membrane organization | 372 | 2.48 | 1.77 | 4.85E-02 |
| GO:0016192 | vesicle-mediated transport | 456 | 3.01 | 2.17 | 3.46E-02 |
| GO:0007242 | intracellular signal transduction | 541 | 3.66 | 2.55 | 6.41E-03 |
| GO:0007389 | pattern specification process | 414 | 2.77 | 1.96 | 3.46E-02 |
| GO:0007399 | nervous system development | 657 | 4.36 | 3.12 | 5.53E-03 |
| GO:0009653 | anatomical structure morphogenesis | 1127 | 7.2 | 5.44 | 1.72E-03 |
| GO:0009887 | organ morphogenesis | 594 | 3.83 | 2.86 | 3.23E-02 |
| GO:0022008 | neurogenesis | 488 | 3.23 | 2.32 | 3.00E-02 |
| GO:0030154 | cell differentiation | 1046 | 6.94 | 4.98 | 2.87E-04 |
| GO:0030182 | neuron differentiation | 433 | 2.89 | 2.05 | 3.28E-02 |
| GO:0009888 | tissue development | 553 | 3.66 | 2.63 | 1.45E-02 |
| GO:0030030 | cell projection organization | 397 | 2.63 | 1.89 | 4.87E-02 |
| GO:0048666 | neuron development | 369 | 2.51 | 1.74 | 3.46E-02 |
| GO:0051726 | regulation of cell cycle | 159 | 1.2 | 0.72 | 4.39E-02 |
| GO:0007610 | behavior | 350 | 2.41 | 1.64 | 3.16E-02 |
| GO:0000278 | mitotic cell cycle | 440 | 2.91 | 2.09 | 3.46E-02 |
| GO:0007422 | peripheral nervous system development | 95 | 0.79 | 0.41 | 4.39E-02 |
| GO:0006163 | purine nucleotide metabolic process | 187 | 1.4 | 0.85 | 3.46E-02 |
| GO:0006164 | purine nucleotide biosynthetic process | 182 | 1.35 | 0.83 | 4.39E-02 |
| GO:0009117 | nucleotide metabolic process | 263 | 1.95 | 1.19 | 9.35E-03 |
| GO:0009165 | nucleotide biosynthetic process | 221 | 1.59 | 1.02 | 4.39E-02 |
| GO:0006836 | neurotransmitter transport | 140 | 1.08 | 0.62 | 4.39E-02 |
| GO:0015672 | monovalent inorganic cation transport | 291 | 2.02 | 1.36 | 3.96E-02 |

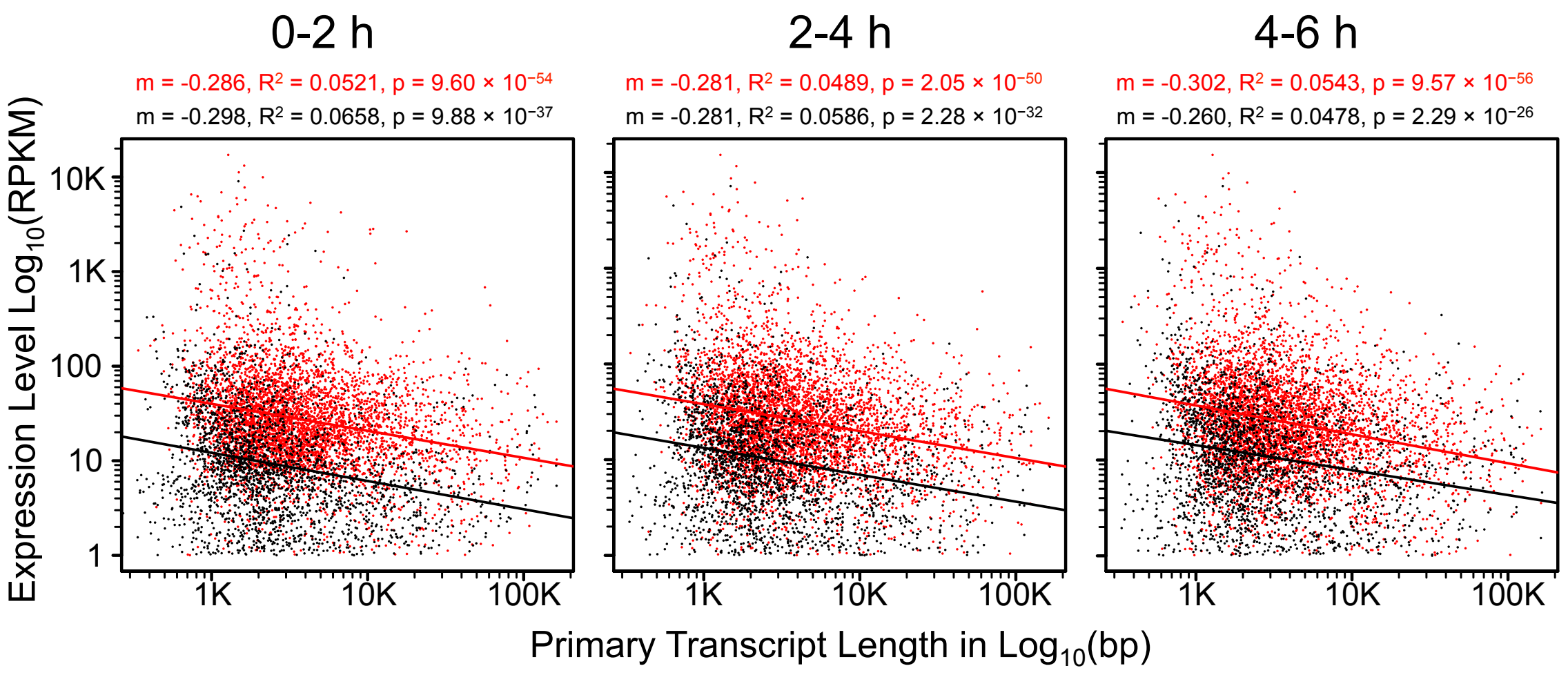

**Figure S1.** Scatterplots of locus length vs. expression level among zygotic (black) and maternal (red) loci for the first three time points of the embryonic timecourse. Slope (m), $R^2$, and p values for the linear regressions for each of the two categories are shown above each time point. The large degree of variance in the data suggests that length explains only a small fraction of total expression level.

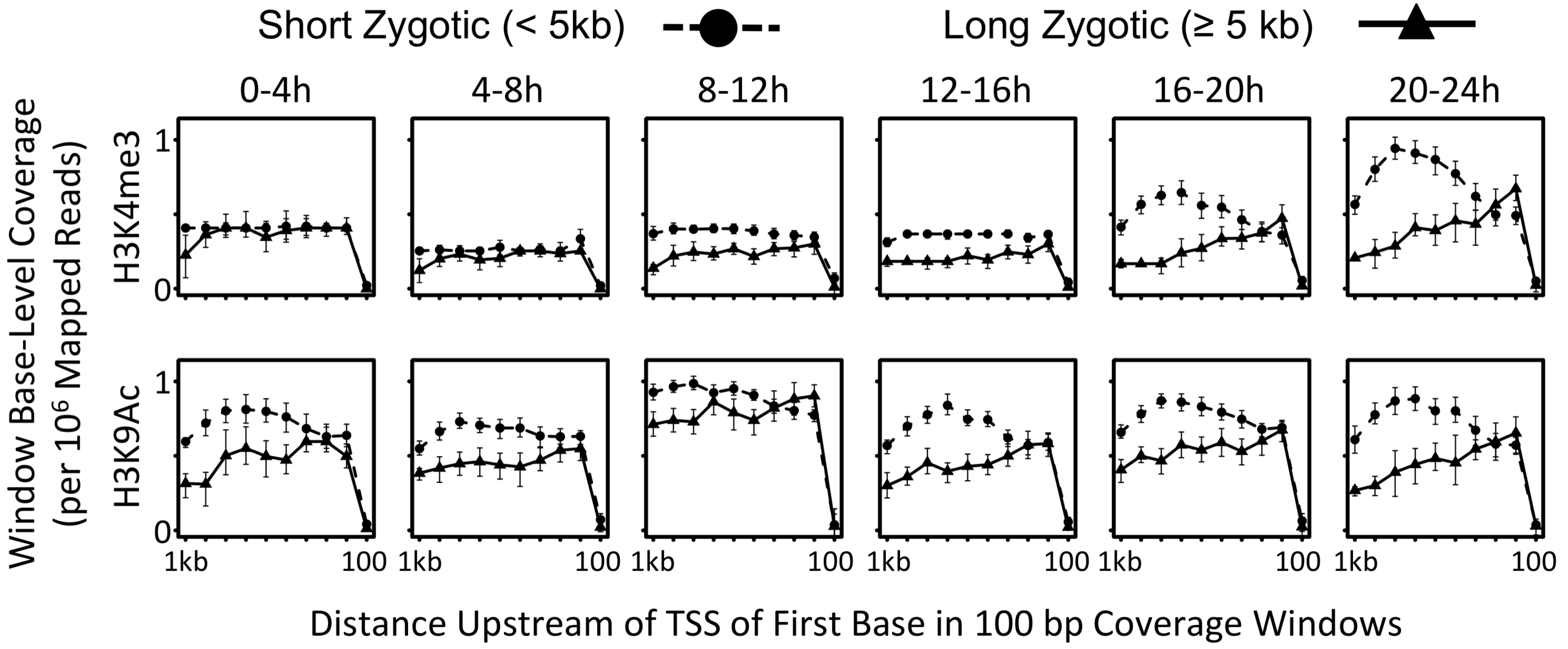

**Figure S2.** Median summed base level coverage (per $10^6$ mapped reads) of euchromatic chromatin marks within 100 bp windows upstream of the TSS of short and long zygotic genes. ChIP-Seq data for transcriptionally activating histone H3 modifications H3K4me3 and H3K9Ac were generated from embryos collected over four-hour windows spanning embryogenesis [25]. Neither mark shows evidence of gradual decrease in the ratio between short and long zygotic genes, as would be expected if the differing patterns of expression of long vs. short zygotic genes were due to differences in transcription initiation rates.

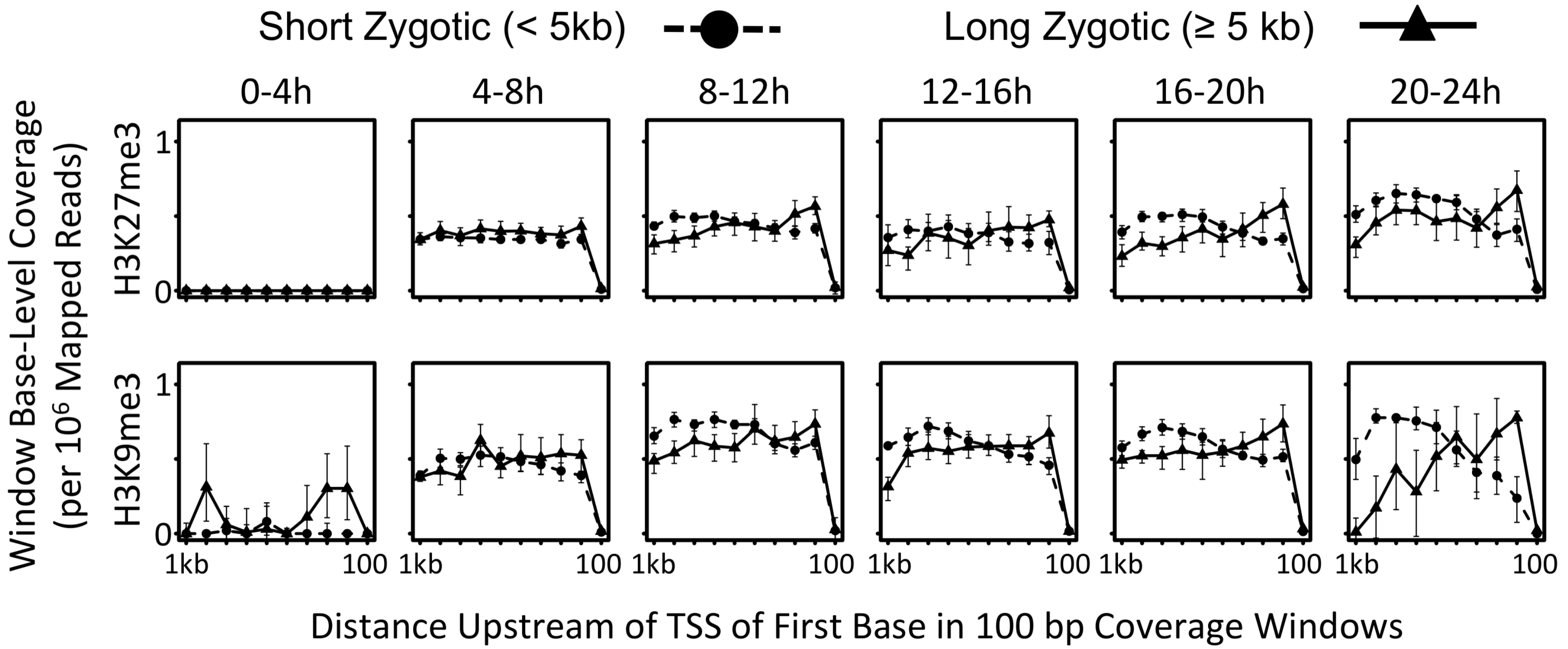

**Figure S3.** Median summed base level coverage (per $10^6$ mapped reads) of heterochromatic chromatin marks within 100 bp windows upstream of the TSS of short and long zygotic genes. ChIP-Seq data for transcriptionally repressive histone H3 modifications H3K27me3 and H3K9me3 were generated from embryos collected over four-hour windows spanning embryogenesis [25]. H3K27me3 does not show evidence of decreasing abundance of repressing chromatin marks spanning development. H3K9me3 shows a significant excess of coverage in the TSSs of long relative to short zygotic genes during the 0-4h time point. However, removal of the long zygotic genes showing increased coverage does not change the conclusions of our analysis (see above).

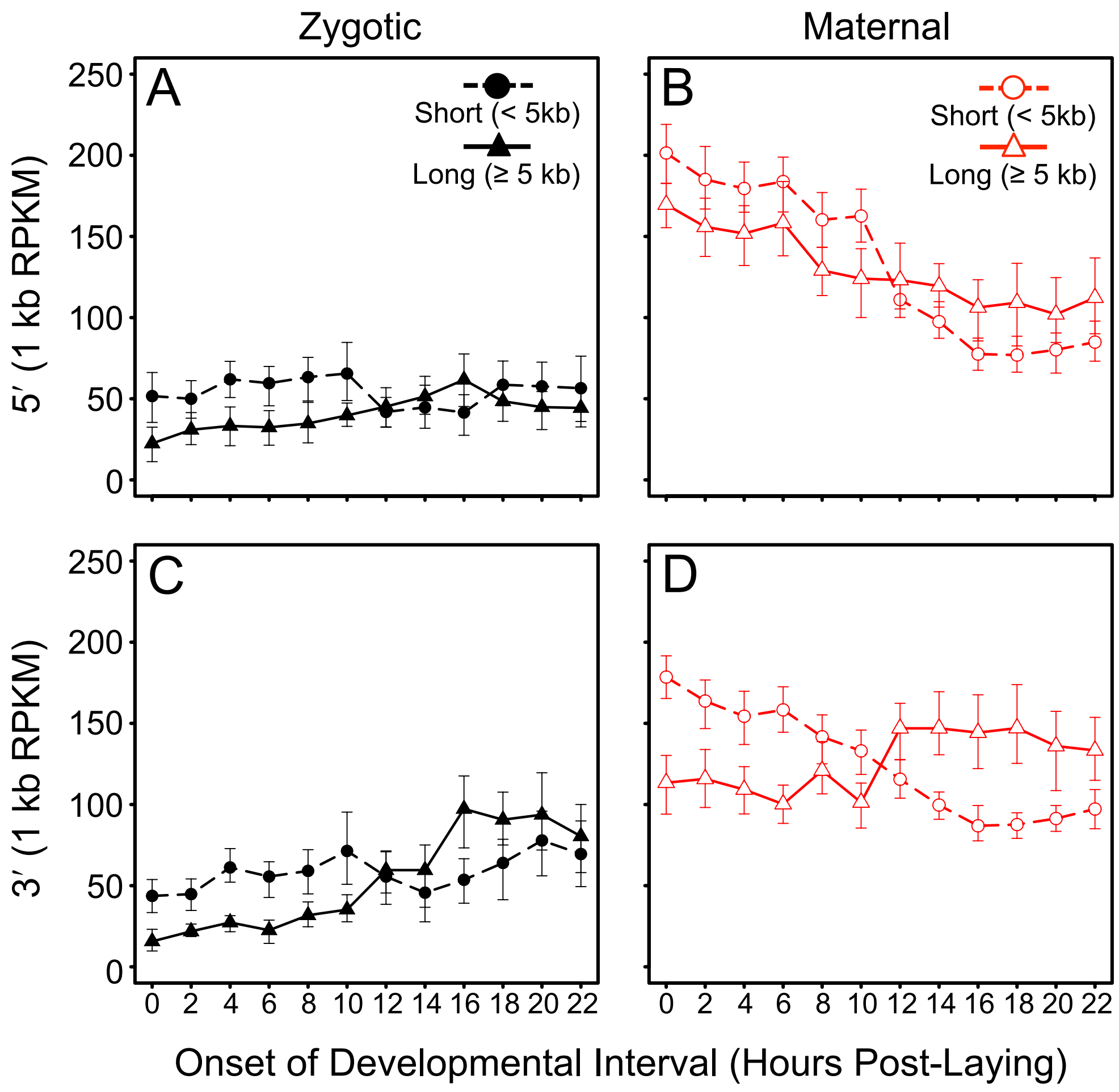

**Figure S4**. Median 5′ and 3′ RPKMs over the embryonic timecourse as determined from total RNA SOLiD data are indicated for zygotic (black) (A, C) and maternal (red) (B, D) loci. Short (< 5 kb) and long (≥ 5 kb) loci are indicated circles and triangles, respectively. Short zygotic loci show relatively modest fluctuations over the timecourse in both 5′ and 3′ RPKMs. Short maternal loci show a general decrease in both transcript ends corresponding to their general decrease in overall expression over embryogenesis (Figure 2B). Long loci in both zygotic and maternal gene categories show an increase in 3′ RPKM in the latter half of embryogenesis.

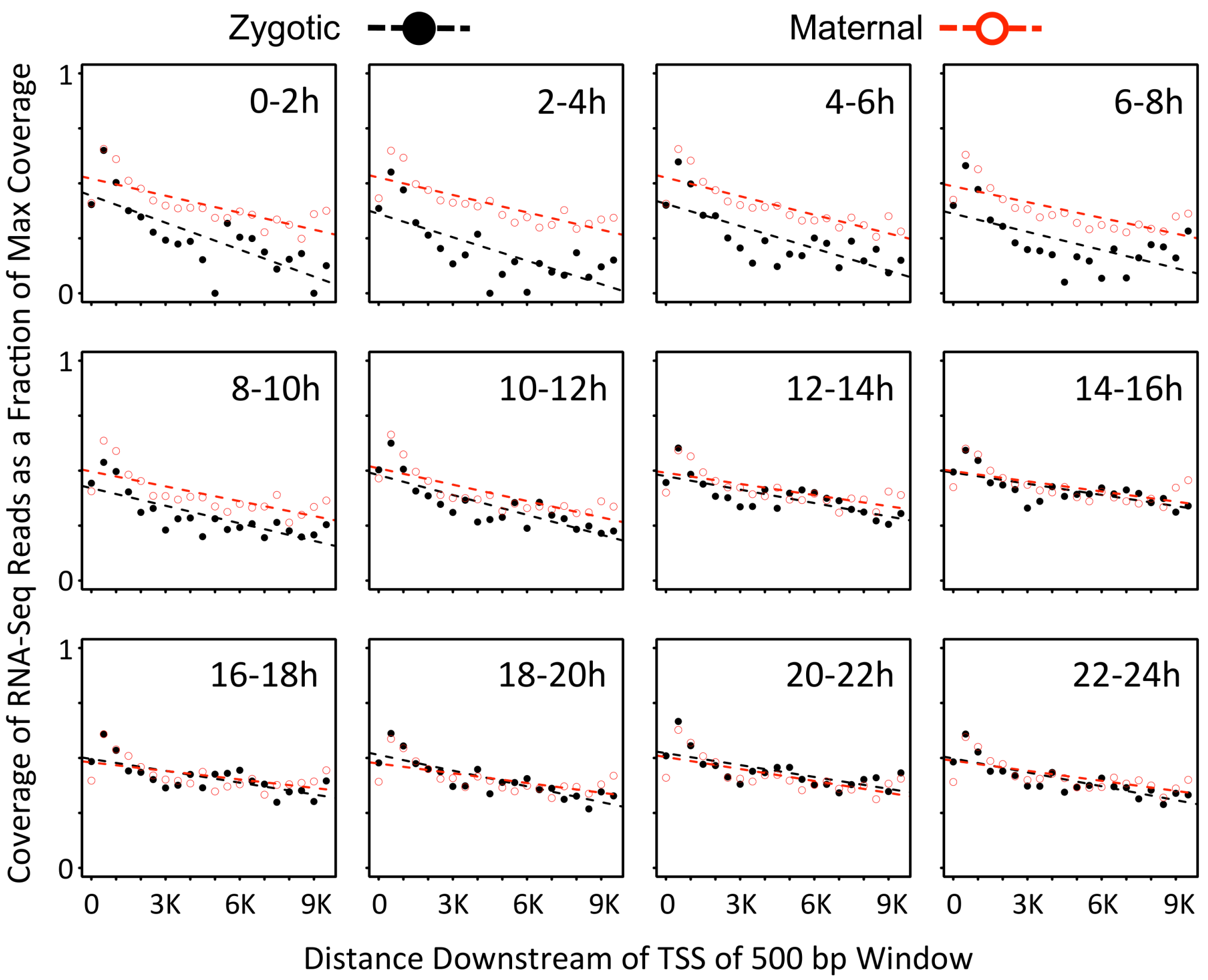

**Figure S5.** Median base-level exonic coverage (normalized as a fraction of maximum coverage) within 500 bp windows over the 5′ most 10 kb of zygotic and maternal transcripts. Both zygotic and maternal transcripts show a slight negative slope, however, the slope is more negative for zygotic as compared to maternal transcripts during the first three time points of development (ANCOVA, $p < 0.05$; note that the ANCOVA for the 0-2h time point is no longer significant after correction for multiple tests is applied). This difference disappears as development progresses, as predicted by the intron delay hypothesis. Note that the increased variability in median coverage in windows further than 5 kb from the TSS among zygotic genes during early development likely reflects the relatively low general expression level of these genes during this period. Only genes with at least 100 mapping reads were included in the analysis during any individual time point.

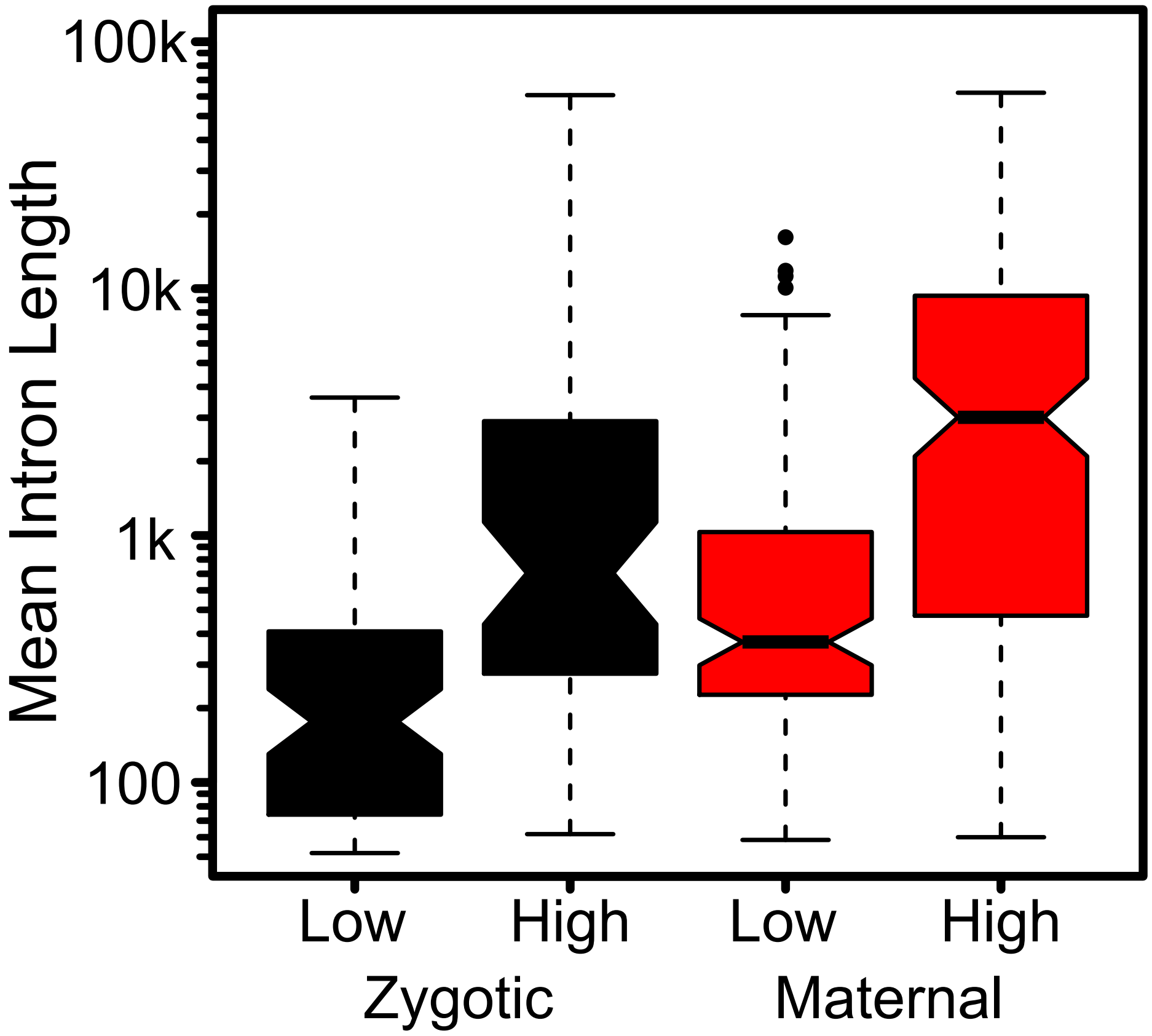

**Figure S6.** Mean orthologous intron length for high and low expression genes during the 0-2 h time point of the four species timecourse among zygotic (black) and maternal (red) genes. The distributions of mean intron length are significantly different among all categories with the exception of the comparison between high expression zygotic and low expression maternal gene categories.